\documentclass[pra,twocolumn,showpacs,preprintnumbers,amsmath,amssymb]{revtex4}
\usepackage{amsfonts}
\usepackage{tipa}

\usepackage{epsfig,graphicx}
\usepackage{amstext}
\usepackage{amsmath}
\usepackage{graphicx}

\begin{document}


\title{Dynamics of entropic measurement-induced nonlocality in structured reservoirs}

\author{Ming-Liang Hu$^{1,}$}
\email{mingliang0301@163.com}
\author{Heng Fan$^{2,}$}
\email{hfan@iphy.ac.cn}
\address{$^{1}$School of Science, Xi'an University of Posts and
               Telecommunications, Xi'an 710061, China \\
         $^{2}$Beijing National Laboratory for Condensed Matter Physics,
               Institute of Physics, Chinese Academy of Sciences, Beijing
               100190, China}

\begin{abstract}
We propose the entropic measurement-induced nonlocality (MIN) as the
maximal increment of von Neumann entropy induced by the locally
non-disturbing measurement, and study behaviors of it both in the
independent and common structured reservoirs. We present schemes for
preserving the MIN, and show that for certain initial states the
MIN, including the quantum correlations, can even be enhanced by the
common reservoir. Additionally, we also show that the different
measures of MIN may give different qualitative characterizations of
nonlocal properties, i.e., it is rather measure dependent than state
dependent.

\end{abstract}


\maketitle

\section{Introduction}
Nonlocality is a basic property of quantum states not available in
classical world, and it has been the research interest for many
years \cite{Bergmann}. The nonlocality can be studied in the context
of Bell-type inequalities \cite{Bell}, it can also be investigated
from other viewpoints, for example, from an information-theoretic
perspective \cite{Brunner}. On the other hand, the nonlocality is
closely related with entanglement which plays a central role in
quantum information processing \cite{Horodeckireview}, but they are
not completely the same. There are a series of work concentrated on
revealing the relationship between nonlocality and entanglement, for
example, it is shown that while there are entangled states without
Bell nonlocality \cite{Masanes}, there also exists nonlocality
without entanglement \cite{Bennett0}.

It is generally accepted that entanglement is a resource in quantum
protocols and algorithms \cite{Horodeckireview}. Interestingly,
recent progress shows that the separable but nonclassical states may
be responsible for the speedup of a quantum algorithm \cite{Knill}.
It ignites interests of people on studying, from various points of
view, the quantumness of correlations
\cite{Ollivier,Luosl,Dakic,Modi} and their relations with
entanglement and nonlocality \cite{Werlang,Cui,Lang,Streltsov}.
Particularly, Luo and Fu \cite{Luof1,Luof2} introduced a geometric
measure of nonlocality which they termed measurement-induced
nonlocality (MIN). In comparison with Bell nonlocality which is
featured by violation of Bell-type inequalities satisfied by any
local hidden variable theory \cite{Nielsen}, the MIN is in some
sense a more general kind of correlation which was introduced from
the consideration that measurements in quantum mechanics usually
causes disturbance. The extent of the maximum disturbance of locally
non-disturbing measurements on the total state can reveal global
effect of that state which cannot be accounted for locally and
therefore can be exploited to quantify nonlocality.

The MIN provides an interesting classification scheme for the
bipartite quantum states. In the original geometric quantification
scheme \cite{Luof2}, it is obtained by maximizing the
Hilbert-Schmidt distance between the pre- and post-measurement
states. For completeness of theoretical consideration, it is also
desirable to quantify it from other perspectives so as to reveal
more about its meaning and properties. This is one of our
motivations in this paper. Moreover, we know that entanglement is
fragile due to decoherence induced by inevitable interaction between
the system and its environment. While the MIN can also be a powerful
resource enabling fascinating quantum tasks as shown in
\cite{Luof2}, it is of practical significance to know the property
of MIN, for example, to find ways to stabilize the MIN of a state or
to minimize the devastating effects of the decohering environments.
This is also a motivation of this work.

In this paper, we introduce the entropic measure of MIN, which is
different from the corresponding geometric measure \cite{Luof2}, and
give an information-theoretic interpretation of it. We will obtain
the conditions for nonvanishing MIN, and show that there is
nonlocality without quantumness in the sense that the MIN may
persist even for certain zero-discord states. We will also offer a
comparative study of the relationships between MIN and different
quantum correlation measures \cite{Ollivier,Luosl,Dakic} for the
two-qubit case. We will show that when interacting with the
structured reservoirs, the MIN can be frozen by introducing detuning
or by performing local unitary operations on the initial state,
particularly, the strength of MIN can even be enhanced by
interacting the system with a common reservoir.

The paper is organized as follows. In Section II we first recall the
concept of MIN and its quantification based on the Hilbert-Schmidt
norm \cite{Luof2}. Then we reformulate the MIN from an entropic way
and list some general features of it. In Section III, we evaluate
dynamics of MIN for a certain family of two-qubit states both in the
independent and common reservoirs, and present schemes for
preserving and enhancing the MIN. Finally, Section IV is devoted to
a summary.

\section{Measures of the MIN}
The preform of MIN was first introduced by Luo and Fu \cite{Luof1}
and further formulated by the same authors \cite{Luof2}. Consider a
bipartite state $\rho$ shared by two parties $a$ and $b$, with
marginal states $\rho^a$ and $\rho^b$, then the MIN is defined as
follows
\begin{equation}
 N^s(\rho)=\max_{\Pi^a}||\rho-\Pi^a(\rho)||^2,
\end{equation}
where the maximum is taken over the local von Neumann measurements
$\Pi^a=\{\Pi_k^a\}$ satisfying $\sum_k \Pi_k^a\rho^a\Pi_k^a=\rho^a$,
and $||\cdot||$ denotes the Hilbert-Schmidt norm with
$||X||^2=\text{tr}X^\dag X$. Clearly, $\Pi^a$ contains in fact only
the measurements that do not disturb $\rho^a$ locally, thus as
pointed out in \cite{Luof2}, the MIN is {\it in some sense dual to}
the geometric measure of quantum discord (GMQD) \cite{Dakic}
$D^s(\rho)=\min_{\Pi^a}||\rho-\Pi^a(\rho)||^2$, with $\Pi^a$ runs
over all the local von Neumann measurements.

By noting that any $m\times n$ bipartite state $\rho$ can always be
represented as
\begin{eqnarray}
\rho&=&\frac{1}{\sqrt{mn}}\frac{\mathbb{I}^a}{\sqrt{m}}
     \otimes\frac{\mathbb{I}^b}{\sqrt{n}}+ \sum_{i=1}^{m^2-1}x_i
     X_i\otimes \frac{\mathbb{I}^b}{\sqrt{n}}\nonumber\\
     &&+\frac{\mathbb{I}^a}{\sqrt{m}}\otimes\sum_{j=1}^{n^2-1}y_j Y_j+
     \sum_{i=1}^{m^2-1}\sum_{j=1}^{n^2-1}r_{ij}X_i\otimes Y_j,
\end{eqnarray}
Luo and Fu derived an analytical formula of MIN for the $2\times n$
dimensional states, which is given by \cite{Luof2}
\begin{equation}
  N^s(\rho)=\left\{
    \begin{aligned}
         &||R||^2-\frac{1}{||\vec{x}||^2}\vec{x}^T R R^T \vec{x}  &\quad\text{if}~\vec{x}\neq 0,\\
         &||R||^2-\lambda_{\text{min}} &\quad\text{if}~\vec{x}=0,
    \end{aligned} \right.
\end{equation}
where $\{X_i\}$ and $\{Y_j\}$ are traceless Hermitian operators
satisfying the conditions $\langle X_i|X_{i'}\rangle=\delta_{ii'}$
and $\langle Y_j|Y_{j'}\rangle=\delta_{jj'}$. Moreover,
$\lambda_{\text{min}}$ is the smallest eigenvalue of $R R^T$ with
$R$ being a real $3\times 3$ matrix with elements $r_{ij}$, and
$\vec{x}=(x_1,x_2,x_3)^T$ with $||\vec{x}||^2=\sum_{i=1}^3 x_i^2$.
Here the superscript $T$ denotes transpose of vectors or matrices.

Although the definitions of MIN and GMQD are different, they may
assume equal values under certain circumstances. To see this,
consider the two-qubit X state $\rho_{\rm X}$ (i.e., the $4\times4$
density matrix with nonzero elements only along the main diagonal
and anti-diagonal), for which one can obtain
\begin{eqnarray}
 &&\vec{x}=\frac{(0,0,\rho_{11}+\rho_{22}-\rho_{33}-\rho_{44})^T}{2},\nonumber\\
 &&\frac{1}{||\vec{x}||^2}\vec{x}^T R R^T \vec{x}=\frac{(\rho_{11}-\rho_{22}-\rho_{33}+\rho_{44})^2}{4},
\end{eqnarray}
and the eigenvalues of $R R^T$ as
\begin{eqnarray}
 &&\lambda_{1,2}=(|\rho_{14}|\pm|\rho_{23}|)^2,\nonumber\\
 &&\lambda_{3}=\frac{(\rho_{11}-\rho_{22}-\rho_{33}+\rho_{44})^2}{4}.
\end{eqnarray}

Moreover, it has been shown by Daki\'{c} {\it et al.} \cite{Dakic}
that for the two-qubit quantum states the GMQD can be presented
explicitly as
\begin{equation}
 D^s(\rho)=\parallel\vec{x}\parallel^2+\parallel
                 R\parallel^2-k_\text{max}.
\end{equation}
where $k_\text{max}$ is the largest eigenvalue of the matrix
$K=\vec{x}\vec{x}^T+RR^T$. For $\rho_{\rm X}$, the eigenvalues of
$K$ can be obtained analytically as
\begin{eqnarray}
 k_{1,2}=\lambda_{1,2},~
 k_{3}=\frac{1}{2}\sum_{n=1}^4 \rho_{nn}^2-(\rho_{11}\rho_{33}+\rho_{22}\rho_{44}).
\end{eqnarray}

Eqs. (4) and (7) imply that
$||\vec{x}||^2+\frac{1}{||\vec{x}||^2}\vec{x}^T R R^T \vec{x}=k_3$,
so for nondegenerate $\rho_{\rm X}^a$ (i.e., $\vec{x}\neq 0$) we
have $N^s(\rho_{\rm X})=D^s(\rho_{\rm X})$ when $k_\text{max}=k_3$.
In fact, the eigenvector of $K$ corresponding to the eigenvalue
$k_3$ is $|e_3\rangle=(0,0,1)^T$, hence the optimal measurement for
obtaining the GMQD is $\Pi_{1,2}^{a}=
\frac{1}{2}(\mathbb{I}\pm\vec{e}\cdot\vec{\sigma})=
\frac{1}{2}(\mathbb{I}\pm\sigma_z)$ \cite{Dakic}. Moreover, if
$\rho_{\rm X}^a$ is nondegenerate, then the unique von Neumann
measurement leaving $\rho_{\rm X}^a$ invariant is
$\Pi_{1,2}^{a}=\frac{1}{2}(\mathbb{I}\pm
\frac{\vec{x}\cdot\vec{\sigma}}{||\vec{x}||})
=\frac{1}{2}(\mathbb{I}\pm\sigma_z)$ \cite{Luof2}. The optimal
measurements for obtaining the MIN and GMQD are completely the same,
thus it is understandable that they assume equal values under the
condition $k_\text{max}=k_3$.

Moreover, it is direct to check that $||\vec{x}||^2-k_3=-\lambda_3$,
thus if $\rho_{\rm X}^a$ is degenerate with the addition of
$k_\text{max}=k_3$ and $\lambda_\text{min}=\lambda_3$, we still have
$N^s(\rho_{\rm X})=D^s(\rho_{\rm X})$.

Besides the Hilbert-Schmidt norm which measures the MIN from a
geometric perspective \cite{Luof2}, it is also desirable to quantify
it from other perspectives. In particular, if we accept that the
quantum mutual information (QMI) is a good measure of total
correlations in a bipartite state $\rho$, then it is natural to
quantify MIN by the maximal discrepancy between the QMI of the pre-
and post-measurement states as
\begin{equation}
 N^v(\rho)=I(\rho)-\min_{\Pi^a}I[\Pi^a(\rho)],
\end{equation}
where $I(\rho)=S(\rho^a)+S(\rho^b)-S(\rho)$ represents the QMI, and
the minimum is taken only over the local von Neumann measurements
leaving $\rho^a$ invariant. This measure of MIN quantifies in fact,
the maximal loss of total correlations under locally invariant
(non-disturbing) measurements. Since
$I[\Pi^a(\rho)]=S(\rho^b)-S(\rho|\{\Pi_k^a\})$ \cite{Ollivier}
represents the average information gained about the system $b$
conditioned on the measurements $\{\Pi^a\}$, $N^v(\rho)$ is in some
sense dual to the quantum discord (QD)
$D^v(\rho)=I(\rho)-\max_{\Pi^a}I[\Pi^a(\rho)]$ \cite{Ollivier}. But
the maximum in $D^v(\rho)$ runs over {\it all} the local von Neumann
measurements. Moreover, the measure of MIN introduced here is also
quite different from the quantity $M(\rho)=I(\rho)-I[\Pi(\rho)]$,
which was introduced by Luo \cite{Luosl} and has been termed
measurement-induced disturbance (MID), because when defining
$M(\rho)$ the locally invariant measurements
$\Pi=\{\Pi^a_i\otimes\Pi^b_j\}$ are performed on both parties of $a$
and $b$.

Since the projective measurement $\Pi^a$ leaves $\rho^a$ invariant
($\rho^b$ is obviously also invariant), $N^v(\rho)$ defined above is
equivalent to
\begin{equation}
 N^v(\rho)=\max_{\Pi^a}S[\Pi^a(\rho)]-S(\rho),
\end{equation}
from which one can obtain $N^v(\rho)\geqslant 0$ because {\it
projective measurements always increase entropy} (Theorem 11.9 of
\cite{Nielsen}). Eq. (9) also indicates that $N^v(\rho)$ quantifies
in fact the maximal increment of von Neumann entropy induced by the
locally invariant measurements. This result thus establishes a
connection between increment of entropy and nonlocality in the
quantum domain. Moreover, from an information-theoretic perspective,
the entropy measures how much uncertainty there is in the state of a
physical system \cite{Nielsen}, $N^v (\rho)$ can therefore also be
interpreted as the maximal increment of our uncertainty about that
system induced by the locally invariant measurements.

The post-measurement state can always be written as
$\Pi^a(\rho)=\sum_{i}p_i|i\rangle\langle i|\otimes\rho^b_i$
\cite{Dakic}, combination of this with the {\it joint entropy
theorem} \cite{Nielsen} gives rise to $S[\Pi^a(\rho)]=H(p_i)+\sum_i
p_i S(\rho^b_i)$, where $H(p_i)$ represents the Shannon entropy.
Furthermore, from Theorem 11.10 of \cite{Nielsen} one can obtain
$H(p_i)+S(\rho^b)\geqslant H(p_i)+\sum_i p_i S(\rho^b_i)\geqslant
S(\rho^b)$, and from \cite{linan} we have $\sum_i p_i
S(\rho^b_i)\leqslant S(\rho)$, thus $N^v(\rho)$ is lower bounded by
$N^v(\rho)\geqslant -S(a|b)$ and upper bounded by
$N^v(\rho)\leqslant{\rm min}\{I(\rho),S(\rho^a)\}$, with
$S(a|b)=S(\rho)-S(\rho^b)$ being the conditional von Neumann entropy
for $\rho$, and $-S(a|b)$ gives the lower bound of the one-way
distillable entanglement. But both the lower and the upper bounds
are non-trivial only when $S(a|b)<0$ because we always have
$N^v(\rho)\geqslant 0$ and $N^v(\rho)\leqslant I(\rho)$.

For the two-qubit X states $\rho_{\rm X}$ with nondegenerate
$\rho_{\rm X}^a$, the only von Neumann measurement leaving
$\rho_{\rm X}^a$ invariant is $\Pi^a=\{\Pi_1^a,\Pi_2^a\}$ with
$\Pi_{1,2}^{a}=\frac{1}{2}(\mathbb{I}\pm\sigma_z)$, which gives rise
to the post-measurement state as $\Pi^a(\rho_{\rm X})=\sum_k
(\Pi_k^a\otimes\mathbb{I})\rho_{\rm X}(\Pi_k^a\otimes\mathbb{I})=
\text{diag}\{\rho_{11},\rho_{22},\rho_{33},\rho_{44}\}$. Still for
$\rho_{\rm X}$ with nondegenerate $\rho_{\rm X}^a$ one can check
directly that $\Pi(\rho_{\rm X})=\sum_{ij}
(\Pi_i^a\otimes\Pi_j^b)\rho_{\rm X}(\Pi_i^a\otimes\Pi_j^b)=
\text{diag}\{\rho_{11},\rho_{22},\rho_{33},\rho_{44}\}$. Thus for
the special circumstance considered here we always have $N^v
(\rho_{\rm X})=M(\rho_{\rm X})$. But if $\rho_{\rm X}^a$ is
degenerate, then $N^v (\rho_{\rm X})\geqslant M(\rho_{\rm X})$ in
general.

Now we discuss conditions under which the MIN disappears. First, it
is clear that the MIN is strictly positive for any state $\rho$ with
non-vanishing quantum correlations. This can be shown directly from
the definitions of MIN and QD based on the Hilbert-Schmidt distance
or the discrepancy between QMI of the pre- and post-measurement
states, for which we always have $N^s(\rho)\geqslant D^s(\rho)$ and
$N^v(\rho)\geqslant D^v(\rho)$. Thus one only need to consider the
classical-quantum state $\rho_{\rm CQ}=\sum_{k}p_k|k\rangle\langle
k|\otimes\rho^b_k$ \cite{Dakic}. If the reduced state $\rho_{\rm
CQ}^a=\sum_{k}p_k|k\rangle\langle k|$ is nondegenerate (i.e.,
$p_k\neq p_l$), then the MIN disappears because for this case the
only von Neumann measurement leaving $\rho_{\rm CQ}^a$ invariant is
$\Pi^a=\{\Pi^a_k=|k\rangle\langle k|\}$, which yields
$\Pi^a(\rho_{\rm CQ})=\rho_{\rm CQ}$ and thus there is no MIN
\cite{Luof2}. If $\rho_{\rm CQ}^a$ is degenerate (i.e., $p_k=p_l$),
then we need to consider two different cases. The first one is that
$\rho^b_k=\rho^b_l$ for all $k$ and $l$. For this case, we always
have $\rho_{\rm CQ}=\rho_{\rm CQ}^a\otimes\rho_{\rm CQ}^b$ and there
is still no MIN. Second, if $\rho^b_k\neq\rho^b_l$, then one can
always find a complete projective measurement
$\Pi^a=\{\Pi^a_{k'}=|k'\rangle\langle k'|\}$ so that
$\Pi^a(\rho_{\rm CQ})\neq\rho_{\rm CQ}$, thus from Theorem 11.9 of
\cite{Nielsen} we obtain $S[\Pi^a(\rho_{\rm CQ})]>S(\rho_{\rm CQ})$,
which gives rise to non-vanishing MIN. Since there is no quantum
correlations in $\rho_{\rm CQ}$, the result obtained here reveals
that there exists not only nonlocality without entanglement but also
nonlocality without quantumness.
\begin{figure}
\centering
\resizebox{0.4\textwidth}{!}{%
\includegraphics{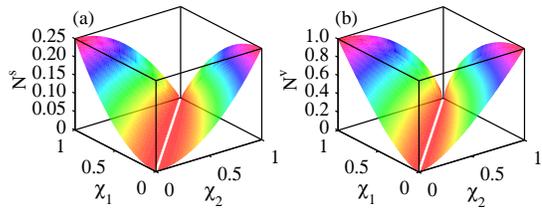}}
\caption{(Color online) MIN versus $\chi_1$ and $\chi_2$ for the
two-qubit $\rho_{\rm CQ}=\sum_{k}p_k|k\rangle\langle
k|\otimes\rho^b_k$ with $p_1=p_2=1/2$,
$\rho^b_k=|\varphi_k\rangle\langle\varphi_k|$, and
$|\varphi_k\rangle=\chi_k|1\rangle+\sqrt{1-\chi_k^2}|0\rangle$.}
\label{fig:1}
\end{figure}

As an example, we consider the two-qubit $\rho_{\rm CQ}$ with
degenerate $\rho_{\rm CQ}^a$ (i.e., $p_1=p_2=1/2$) and
$\rho^b_k=|\varphi_k\rangle\langle\varphi_k|$. In Fig. 1 we plot
$N^s(\rho_{\rm CQ})$ and $N^v(\rho_{\rm CQ})$ versus $\chi_1$ and
$\chi_2$ with
$|\varphi_k\rangle=\chi_k|1\rangle+\sqrt{1-\chi_k^2}|0\rangle$. It
is evident that the MIN vanishes only when $\chi_1=\chi_2$, that is
$\rho^b_1=\rho^b_2$, which corroborates our above statements. Also
one can note that when $\chi_1=1$, $\chi_2=0$ or $\chi_1=0$,
$\chi_2=1$, $N^s(\rho_{\rm CQ})$ assumes the value 0.25, half of the
maximum 0.5 obtained for the maximally entangled Bell states
$\rho_{\rm Bell}$. But $N^v(\rho_{\rm CQ})$ assumes its maximum 1
under the same conditions, equals completely to those obtained for
$\rho_{\rm Bell}$. This is because $S(\rho_{\rm CQ})=1$, and for
this two special cases one can always choose corresponding complete
projective measurements $\Pi^a=\{\Pi^a_1,\Pi^a_2\}$ so that
$\Pi^a(\rho_{\rm CQ})=\mathbb{I}/4$ and $S[\Pi^a(\rho_{\rm CQ})]=2$.
In fact, the classical correlation in $\rho_{\rm CQ}$ is completely
destroyed by the projective measurement $\Pi^a$ because
$I[\Pi^a(\rho_{\rm CQ})]=0$. For $\rho_{\rm Bell}$, however, the
classical correlation is undisturbed because the optimal measurement
$\Pi^a$ (may be different from that for $\rho_{\rm CQ}$) for
obtaining the MIN gives $I[\Pi^a(\rho_{\rm Bell})]=1$, equals to the
classical correlation present in $\rho_{\rm Bell}$ \cite{Ollivier}.
Moreover, we would like to point out that for $\rho_{\rm CQ}$ with
non-vanishing MIN, the classical correlation is partially destroyed
by the optimal measurement $\Pi^a$.

The above discussions also indicate that the different measures of
MIN may impose different orderings of quantum states, and in
particular, the maximum $N^v(\rho)=1$ does not always indicates that
the state $\rho$ is maximally entangled, which is in sharp contrast
to that of MIN measured by $N^s(\rho)$ or quantum correlations
measured by QD, GMQD and MID \cite{Ollivier,Dakic,Luosl}.

\section{MIN dynamics in structured reservoirs}
\begin{center}
\textbf{A. The model}
\end{center}

We consider a system consists of two identical atoms with lower and
upper levels denoted by $|0\rangle$ and $|1\rangle$, and is coupled
to two typical structured reservoirs, i.e., the independent and
common zero-temperature bosonic reservoir \cite{Breuer}. For the
former case, the two atoms interact with their own reservoir with
$H=H_1+H_2$, and the single ``qubit+reservoir'' Hamiltonian reads
\begin{equation}
 H_n=\omega_0\sigma^n_{+}\sigma^n_{-}
   +\sum_{k}\omega_k^n b_k^{n\dag} b_k^n+\sum_{k}(g_k^n b_k^n\sigma^n_{+}+h.c.),
\end{equation}
where $\omega_0$ denotes the transition frequency of the atoms, and
$\sigma^n_\pm$ ($n=a,b$) are the Pauli rasing and lowering
operators. The index $k$ labels the reservoir field mode with
frequency $\omega_k^n$, with $b_k^{n\dag}$ and $b_k^n$ being the
bosonic creation and annihilation operators and $g_k^n$ the coupling
strength. We will assume that the two reservoirs are completely the
same and denote $\omega_k^{1,2}\equiv\omega_k$ and $g_k^{1,2}\equiv
g_k$.

For the latter case, the two atoms are coupled to the same reservoir
and the Hamiltonian reads
\begin{equation}
 H=\omega_0\sum_n\sigma^n_{+}\sigma^n_{-}
   +\sum_{k}\omega_k b_k^{\dag} b_k+\sum_{k,n}(g_k b_k
   \sigma^n_{+}+h.c.).
\end{equation}
In the following, we consider the structured reservoir with the
spectral density having the Lorentzian form \cite{Breuer}
\begin{equation}
 J(\omega)=\frac{1}{2\pi}\frac{\gamma_0\lambda^2}{(\omega-\omega_c)^2+\lambda^2},
\end{equation}
where $\lambda$ denotes the spectral width of the reservoir and is
related to the reservoir correlation time $\tau_B$ by the relation
$\tau_B\approx\lambda^{-1}$, and $\gamma_0$ is related to the decay
time scale $\tau_R$ of the excited state of the atom in the
Markovian limit of flat spectrum via $\tau_R\approx\gamma_0^{-1}$.
$\lambda>2\gamma_0$ corresponds to the Markovian regime, while
$\lambda<2\gamma_0$ corresponds to the non-Markovian regime.
Moreover, $\omega_c=\omega_0-\delta$ is the central frequency of the
reservoir detuned from the transition frequency $\omega_0$ by an
amount $\delta$.

For the independent reservoir, we work in the interaction picture
and thus the Hamiltonian of Eq. (10) can be rewritten as
$H_{n,I}(t)=\sum_k g_k\sigma^n_{+}e^{i(\omega_0-\omega_k)t}+h.c.$.
We suppose the initial single ``qubit+reservoir'' state to be
$|\psi(0)\rangle=c_0|0\bar{0}\rangle+c_1(0)|1\bar{0}\rangle$
($|\bar{0}\rangle$ represents vacuum state of the reservoir), then
by using the Schr\"{o}dinger equation $i \partial
|\psi(t)\rangle/\partial t=H_{n,I}(t)|\psi(t)\rangle$ with
$|\psi(t)\rangle=c_0|0\bar{0}\rangle+c_1(t)|1\bar{0}\rangle
+\sum_kc_k(t)|0\bar{1}_k\rangle$ ($|\bar{1}_k\rangle$ represents the
reservoir with one excitation in mode $k$), one can obtain
\begin{eqnarray}
 &&\dot{c}_1(t)=-i\sum_k g_k c_k(t)e^{i(\omega_0-\omega_k)t},\nonumber\\
 &&\dot{c}_k(t)=-ig_k^* c_1(t)e^{-i(\omega_0-\omega_k)t}.
\end{eqnarray}

From the second expression of the above equation one can derive
$c_k(t)=-ig_k^*\int_0^tdt_1 c_1(t_1)e^{-i(\omega_0-\omega_k)t_1}$,
substituting of which into the first expression we obtain
\begin{equation}
 \dot{c}_1(t)=-i\int_0^tdt_1f(t-t_1)c_1(t_1).
\end{equation}
where the function $f(t-t_1)=\int d\omega
J(\omega)e^{-i(\omega_0-\omega)(t-t_1)}=\frac{\gamma_0 \lambda}{2}
e^{-(\lambda-i\delta)(t-t_1)}$ in the limit of large number of modes
$k$. Then by performing the Laplace transformation of Eq. (14) one
can obtain $c_1(t)=p(t)c_1(0)$ with
\begin{equation}
 p(t)=e^{-\frac{1}{2}(\lambda-i\delta)t}\left[\cosh\frac{dt}{2}
 +\frac{\lambda-i\delta}{d}\sinh\frac{dt}{2}\right],
\end{equation}
where $d=\sqrt{(\lambda-i\delta)^2-2\gamma_0\lambda}$. If we further
define $|\bar{1}\rangle=c^{-1}(t)\sum_k c_k(t)|\bar{1}_k\rangle$,
then $|\psi(t)\rangle=c_0|0\bar{0}\rangle+c_1(t)|1\bar{0}\rangle
+c(t)|0\bar{1}\rangle$, according to which the single-qubit reduced
density matrix can be determined as
\begin{equation}
 \rho^{S}(t)=\left(\begin{array}{cc}
  \rho_{11}^S(0)|p(t)|^2  & \rho_{10}^S(0)p(t) \\
  \\
  \rho_{01}^S(0)p^*(t) & 1-\rho_{11}^S(0)|p(t)|^2
 \end{array}\right),
\end{equation}
in the standard basis $\{|1\rangle,|0\rangle\}$. For the two-qubit
system considered in this work its reduced density matrix can be
obtained by the procedure presented in Ref. \cite{Bellomo1}.

If the two atoms are coupled to the common reservoir, then their
dynamics can be described by the following pseudomode master
equation \cite{Garraway,Mazzola1,Mazzola2,Rossato}
\begin{equation}
 \frac{\partial\tilde{\rho}}{\partial t}=-i[V,\tilde{\rho}]+
 \lambda(2b\tilde{\rho}
 b^{\dag}-b^{\dag}b\tilde{\rho}-\tilde{\rho} b^{\dag}b),
\end{equation}
where $\tilde{\rho}$ is the density matrix for the atoms and the
pseudomode, and the interaction Hamiltonian $V$ of the atomic system
and the pseudomode can be written as
\begin{equation}
 V=\sqrt{\frac{\gamma_0\lambda}{2}}(\sigma^a_{+}+\sigma^b_{+})b+h.c.,
\end{equation}

We assume throughout this work that the total system contains at
most two excitations, then Eq. (17) turns out to be set of 64
differential equations, which can be solved by using different
methods \cite{Maniscalco,Mazzola1,Mazzola2,Rossato,epjd}.
Particularly, for the initial extended Werner-like (EWL) states,
analytical solutions of Eq. (17) were obtained by using the Laplace
transform which turns the differential equations to the polynomial
equations \cite{Mazzola2} (note that there may be some misprints in
\cite{Mazzola2}. First, the $12\Omega^2$ in Eq. (A.21) should be
$12\Omega^2 s$, and $\alpha r\sqrt{1-\alpha^2}$ should be multiplied
by $e^{-i\theta}$. Second, the solution of $\tilde{\rho}_{--}(t)$ in
Eq. (A.22) should be $\tilde{\rho}_{--}(t)=\frac{1-r}{4}$).
\\

\begin{center}
\textbf{B. MIN dynamics}
\end{center}

Now we begin our discussion about MIN dynamics in structured
reservoirs. We suppose the two atoms are prepared initially in the
EWL states
\begin{eqnarray}
 &&\rho_0^{\Xi}=r|\Xi\rangle\langle\Xi|+\frac{1-r}{4}\mathbb{I},
\end{eqnarray}
with $|\Xi\rangle=|\Psi\rangle$ or $|\Phi\rangle$, and
$|\Psi\rangle=\alpha|00\rangle+e^{i\theta}\sqrt{1-\alpha^2}|11\rangle$,
$|\Phi\rangle=\alpha|10\rangle+e^{i\theta}\sqrt{1-\alpha^2}|01\rangle$.

\begin{figure}
\centering
\resizebox{0.4\textwidth}{!}{%
\includegraphics{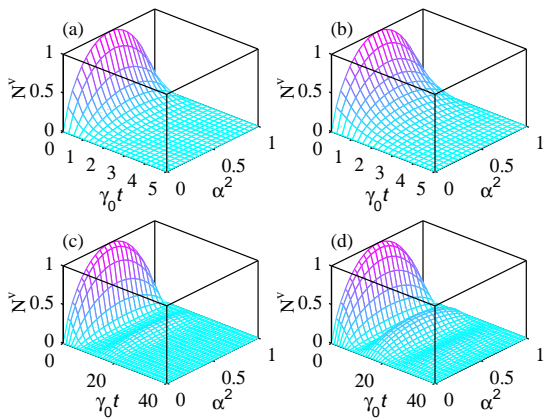}}
\caption{(Color online) $N^v(\rho)$ versus $\alpha^2$ and $\gamma_0
t$ for different initial states in independent reservoirs. (a)
$|\Psi\rangle$ with $\lambda=10\gamma_0$, (b) $|\Phi\rangle$ with
$\lambda=10\gamma_0$, (c) $|\Psi\rangle$ with $\lambda=0.1\gamma_0$,
and (d) $|\Phi\rangle$ with $\lambda=0.1\gamma_0$. $N^s(\rho)$ shows
qualitatively similar behaviors and hence we have not presented them
here.} \label{fig:2}
\end{figure}

For the case of independent reservoir, $\rho_0^\Psi$ and
$\rho_0^\Phi$ yield qualitatively similar MIN dynamics, which are
independent of the phase factor $e^{i\theta}$. In Fig. 2 we plotted
$N^v(\rho)$ versus $\alpha^2$ and $\gamma_0 t$ with the two atoms
prepared in different initial states and coupled resonantly (i.e.,
$\delta=0$) to the reservoirs. It is clear that they attain certain
maxima at $\alpha^2=0.5$ and vanish at $\alpha^2=0$ or $1$. For
fixed $\alpha^2$, MIN decreases exponentially with increasing
$\gamma_0 t$ for the Markovian reservoir [see Figs. 2(a) and 2(b)],
while there are revivals of MIN for the non-Markovian reservoir [see
Figs. 2(c) and 2(d)]. Since there are neither direct nor
reservoir-mediated interactions between the two atoms, the revivals
of MIN after a certain minimum may be induced by the non-Markovian
effects.

\begin{figure}
\centering
\resizebox{0.4\textwidth}{!}{%
\includegraphics{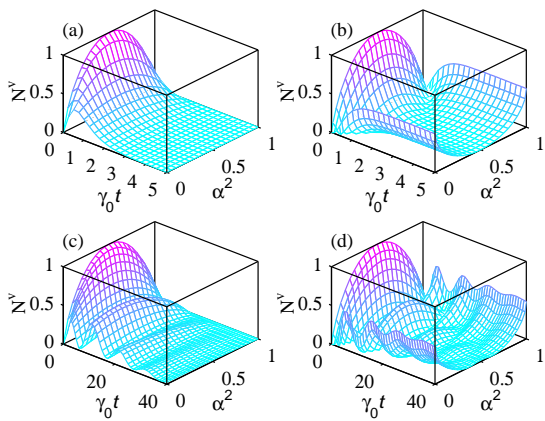}}
\caption{(Color online) $N^v(\rho)$ versus $\alpha^2$ and $\gamma_0
t$ for different initial states in common reservoir. (a)
$|\Psi\rangle$ with $\lambda=10\gamma_0$, (b) $|\Phi\rangle$ with
$\lambda=10\gamma_0$, $\theta=0$, (c) $|\Psi\rangle$ with
$\lambda=0.1\gamma_0$, and (d) $|\Phi\rangle$ with
$\lambda=0.1\gamma_0$, $\theta=0$. $N^s(\rho)$ shows qualitatively
similar behaviors and hence we have not presented them here.}
\label{fig:3}
\end{figure}

If the two atoms interact with the common reservoir, however,
$\rho_0^\Psi$ and $\rho_0^\Phi$ will exhibit completely different
MIN dynamics. Particularly, for the initial sub-radiant state
$|\Psi^{-}\rangle=(|10\rangle-|01\rangle)/\sqrt{2}$, the density
matrix maintains its initial value (i.e., $\rho(t)\equiv
|\Psi^{-}\rangle\langle\Psi^{-}|$) \cite{Mazzola1} and thus the MIN
does not decay. To see more about MIN dynamics for this case, we
show in Fig. 3 the $N^v(\rho)$ versus $\alpha^2$ and $\gamma_0 t$
with different initial states. One can note that the MIN displays
complicated behaviors. First, it is not a symmetric quantity with
respect to $\alpha^2=0.5$ for the initial state $|\Psi\rangle$, and
$N^v(\rho)$ can even does not behave as a monotonic function of
$\gamma_0 t$ for the Markovian reservoir. As one can see from Figs.
3(a) and 3(b), there are creations of MIN for the initial state
$|\Psi\rangle$ with $\alpha^2=0$ and for $|\Phi\rangle$ with
$\alpha^2=0$ or $1$. The underlying reason for this is the
reservoir-mediated interaction because here the two atoms interact
with a common reservoir. Second, there are still revivals of MIN for
the non-Markovian case, but the combined effects of the
non-Markovian effects and the reservoir-mediated interaction makes
it more evident than those for the independent reservoirs.

\begin{figure}
\centering
\resizebox{0.4\textwidth}{!}{%
\includegraphics{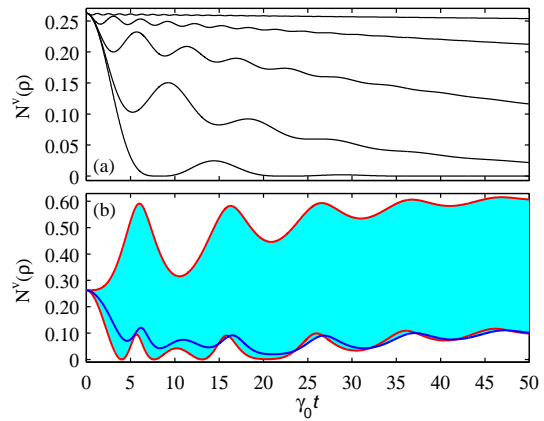}}
\caption{(Color online) (a) $N^v(\rho)$ versus $\gamma_0 t$ for the
initial state $\rho_0^{\Psi}$ in independent reservoir. The curves
from bottom to top correspond to $\delta/\gamma_0=0,0.5,1,2,5$. (b)
$N^v(\rho)$ versus $\gamma_0 t$ for the initial states
$\rho_0^{\Psi}$ (blue), $\rho_0^{\Phi}$ with $\theta=\pi$ (red, top)
or $\theta=0$ (red, bottom) in common reservoir. The cyan shaded
region represents the MIN that are achievable by performing local
unitary operations on the qubits. The other parameters for the plots
are $\alpha^2=1/2$, $r=1/2$ and $\lambda=0.1\gamma_0$.}
\label{fig:4}
\end{figure}

When the two atoms interact with independent reservoirs, their
entanglement may be trapped by detuning the central frequency of the
reservoir from $\omega_0$ \cite{Bellomo2}. In Fig. 4(a) we display
$N^v(\rho)$ versus $\gamma_0 t$ for the initial state
$\rho_0^{\Psi}$ with various values of detuning $\delta$. The result
shows that $N^v(\rho)$ may be enhanced by introducing detuning
$\delta$ and when $\delta=5\gamma_0$ it begins to oscillating weakly
around its initial value, from which it is reasonable to conjecture
that up to the large detuning limit case $N^v(\rho)$ will maintains
its initial value during the time evolution process and thus MIN
trapping occurs. Experimentally, for two Rb Rydberg atoms with
lifetime $T_{\rm at}\approx 30~{\rm ms}$ and inside the
Fabry-P\'{e}rot superconducting resonant cavities with quality
factor $Q\approx 4.2\times 10^{10}$, we have the atomic decay rate
$\gamma_0\approx 33.3~{\rm Hz}$, and the spectral width of the
reservoir $\lambda\approx 7~{\rm Hz}$ \cite{Kuhr}. These values
correspond to a good non-Markovian regime. Moreover, the detuning
needed to freeze the MIN can be implemented by Stark-shifting the
transition frequencies with a static electric field. The typical
Stark shifts are $\delta\sim 200~{\rm kHZ}$ \cite{Hagley} and thus a
dimensionless shift with respect to $\gamma_0$ of the order
$\delta/\gamma_0\sim 6\times 10^3$ can be realized.

Since the reservoir always acts in some basis and therefore changing
the basis may change the strength of reservoir and thus affect the
durability of correlations. Particularly, when the two identical
atoms equally coupled to the common reservoir, there exists a
sub-radiant state $|\Psi^{-}\rangle$ which is decoupled from the
field modes and thus does not decay with time \cite{Mazzola1}. This
motivates us to enhance robustness of MIN by performing appropriate
local unitary operations on the initial state, though the MIN itself
is invariant under local unitary operations. As an explicit example,
we show in Fig. 4(b) the numerical results obtained from the initial
local unitary equivalent states $(U\otimes V)\rho_0^\Psi
(U^\dag\otimes V^\dag)$ with $\alpha^2=1/2$, $r=1/2$ and
$\lambda=0.1\gamma_0$. It is clear that the strength of MIN at
finite time $t> 0$ can be tuned among regions bounded approximately
by those obtained for the initial states $\rho_0^\Phi$ with
$\theta=0$ and $\theta=\pi$, respectively. Particularly, one can
note that for the initial state $\rho_0^\Phi$ with $\theta=\pi$ (the
top red line), the amount of MIN can even be enhanced by subjecting
the two atoms to a common reservoir.

\begin{figure}
\centering
\resizebox{0.4\textwidth}{!}{%
\includegraphics{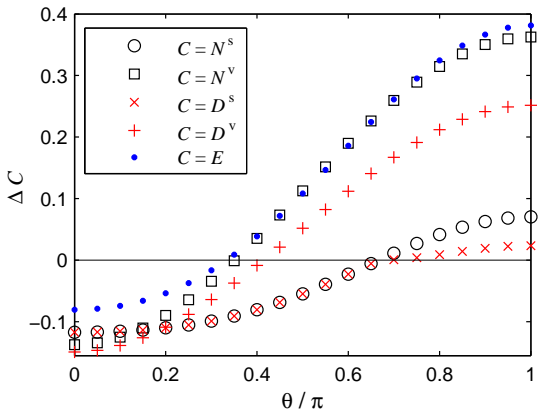}}
\caption{(Color online) Numerical results of $\Delta C=C(t)-C(0)$
versus $\theta/\pi$ for the initial state $\rho_0^{\Phi}$ with
$\alpha^2=1/2$, $r=1/2$ , $\lambda=0.1\gamma_0$ and
$t=5000/\gamma_0$.} \label{fig:5}
\end{figure}

In fact, for the initial state $\rho_0^\Phi$ there exists a
threshold phase angle $\theta_{\rm th}$ (depends on the system
parameters) beyond which the MIN as well as the quantum correlations
measured by entanglement of formation (denoted by $E$)
\cite{Bennett}, QD and GMQD can be enhanced to a larger value and
further be frozen in the long time limit. In Fig. 5 we display
numerical results of $\Delta C=C(t)-C(0)$ versus $\theta/\pi$ for
$\rho_0^{\Phi}$ with $\alpha^2=1/2$, $r=1/2$, $\lambda=0.1\gamma_0$
and $t=5000/\gamma_0$. Clearly, the threshold $\theta_{\rm th}$ are
different for different correlation measures, but $\Delta C$ always
increases with increasing value of $\theta$ and attains the maxima
at $\theta=\pi$. This provides an efficient method for enhancing and
storing MIN for future usage in external environments. One can also
note that the curves for $N^s(\rho)$ and $D^s(\rho)$ are overlapped
when $\theta<\theta_{\rm th}$, this is because during this parameter
region $\rho^a$ is nondegenerate and $k_3$ is the largest eigenvalue
of $K$ (see Section II).

\begin{figure}
\centering
\resizebox{0.4\textwidth}{!}{%
\includegraphics{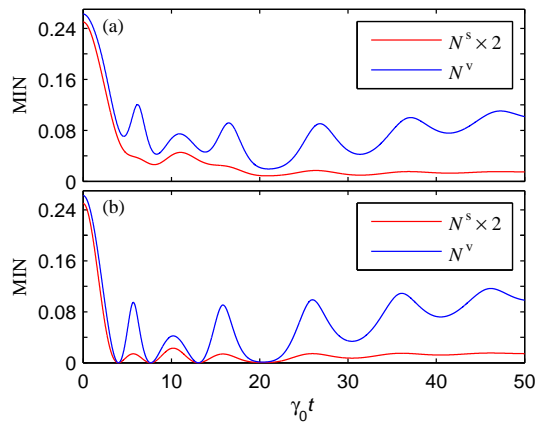}}
\caption{(Color online) Comparison of the geometric and entropic MIN
in common reservoir with $\alpha^2=1/2$, $r=1/2$ and
$\lambda=0.1\gamma_0$. The plot in (a) shows the case of the initial
state $\rho_0^{\Psi}$, while (b) shows that of $\rho_0^{\Phi}$ with
$\theta=0$.} \label{fig:6}
\end{figure}

Finally, we compare dynamical behaviors of MIN quantified by
different measures. Two exemplified plots with MIN quantified by
$N^s(\rho)$ and $N^v(\rho)$ are presented in Fig. 6, from which one
can note that while for the initial state $\rho_0^\Phi$ the two
different measures give qualitatively similar characterizations of
the MIN, for $\rho_0^\Psi$ there exists states $\rho(t_1)$ and
$\rho(t_2)$ which fulfill $N^s[\rho(t_1)]<N^s[\rho(t_2)]$ and
$N^v[\rho(t_1)]>N^v[\rho(t_2)]$. Thus we see that the different
measures of MIN are not only quantitatively but also qualitatively
different in the sense that they may impose inequivalent orderings
of quantum states. Therefore, the lack of the same states ordering
under different entanglement measures \cite{Miranowica} and other
quantum correlation measures \cite{Yeo}, goes beyond entanglement
and quantum correlations.

\section{Summary}
In this paper, we introduced the entropic measure of MIN based on
the maximal difference between QMI of the pre- and post-measurement
states. While the geometric measure of MIN \cite{Luof2} quantifies
the maximal Hilbert-Schmidt distance between $\rho$ and
$\Pi^a(\rho)$, the entropic measure of MIN introduced here
quantifies the maximal loss of total correlations or equivalently,
the maximal increment of von Neumann entropy of $\rho$ by $\Pi^a$.
It seems naturally a definition from information-theoretical point
of view. Based on this measure, we discussed conditions for
nonvanishing MIN and consolidated the finding that there exists
nonlocality without quantumness \cite{Luosl}. For the two-qubit
case, we further compared MIN with different quantum correlation
measures \cite{ Ollivier,Luosl,Dakic}, showing that some of them may
assume equal values under certain special conditions, and a quantum
state with the maximum $N^v(\rho)=1$ does not always indicate that
the state possess the maximum quantum correlation.

Since real physical system is inevitably coupled to its
surroundings, in comparing with fragile entanglement, it is
significant to study stability of MIN against decoherence. Here as a
concrete example, we evaluated MIN dynamics for a family of
two-qubit states in the Lorentzian structured reservoirs, and
displayed their rich dynamical behaviors both for the Markovian and
non-Markovian situations. We showed that the MIN can be frozen by
increasing the detuning parameter, and for some initial states the
strength of MIN as well as the quantum correlations can even be
enhanced in the short time region and further be frozen in the long
time limit by interacting the system with a common reservoir, the
underlying reason for which may be the existent component of the
decoherence-free state $|\Psi^{-}\rangle$.

Finally, we studied the issue related to the relativity of different
MIN measures, just as that of different entanglement
\cite{Miranowica} and quantum correlation measures \cite{Yeo}. By
evaluating the dynamics of $N^s(\rho)$ and $N^v(\rho)$ in an
non-Markovian common reservoir, we showed that the two MIN measures
do not necessarily imply the same ordering of quantum states. This
is reminiscent of the cases for entanglement and quantum correlation
measures and may indicate that the difference in nonlocality is
rather measure dependent than state dependent. Experimentally,
besides atomic systems \cite{Bellomo2,Kuhr,Hagley}, two-qubit
quantum state with different reservoirs, such as Markovian and
non-Markovian, can be realized \cite{guonaturep}, and different
correlations such as quantum discord have been shown recently
\cite{guonaturec}. It will also be interesting to explore the
entropic MIN of this paper in experiments.
\\

\begin{center}
\textbf{ACKNOWLEDGMENTS}
\end{center}

This work was supported by NSFC (10974247, 11175248), ``973''
program (2010CB922904), NSF of Shaanxi Province (2010JM1011,
2009JQ8006), and the Scientific Research Program of Education
Department of Shaanxi Provincial Government (2010JK843).

\newcommand{\PRL}{Phys. Rev. Lett. }
\newcommand{\PRA}{Phys. Rev. A }
\newcommand{\JPA}{J. Phys. A }
\newcommand{\JPB}{J. Phys. B }
\newcommand{\PLA}{Phys. Lett. A }
%

%

\end{document}